\documentclass{aa}

\newcommand{\kms}{{\,km\,s$^{-1}$}}

\newcommand{\vsini}{{$v\sin i$}}

\begin{document}
   \title{The first close-up of the ``flip-flop'' phenomenon in a single 
star\thanks{Based on the observations obtained at the Kitt Peak National 
Observatory, USA; the Automatic Photometric Telescope, Phoenix 10, Arizona, 
USA; the Nordic Optical Telescope, Observatorio Roque de los Muchachos, 
La Palma, Canary Islands, Spain.}} 

   \author{H. Korhonen \inst{1}
   \and S.V. Berdyugina \inst{1}
   \and K.G. Strassmeier \inst{2}
   \and I. Tuominen \inst{1}}

   \offprints{H.\ Korhonen: heidi.korhonen@oulu.fi}

\institute{ Astronomy Division, P.O. Box 3000, FIN-90014 University of Oulu, 
Finland
\and Astrophysikalisches Institut Potsdam, An der Sternwarte 16, D-14482 
Potsdam, Germany}

   \date{Received; accepted }

\abstract{
We present temperature maps of the active late-type giant FK~Com which exhibit
the first imagining record of the ``flip-flop'' phenomenon in a single star. 
The phenomenon, in which the main part of the spot activity 
shifts 180\degr\ in longitude, discovered a decade ago in FK~Com, was reported 
later also in a number of RS~CVn binaries and a single young dwarf. 
With the surface images obtained right before and after the ``flip-flop'',
we clearly show that the ``flip-flop'' phenomenon in FK~Com is caused by 
changing the relative strengths of the spot groups at the two active 
longitudes, with no actual spot movements across the stellar surface, 
i.e.\ exactly as it happens in other active stars.
}

   \maketitle

\section{Introduction} 

In order to study the spot evolution in cool active stars, two approaches are 
commonly used: photometric light curve modeling (or inversion) and Doppler 
imaging. The light curve analysis is evidently a much poorer source of 
information than Doppler imaging and, therefore, leaves more freedom for 
conclusions about the spot evolution. However, the amount, frequency and time 
scale of photometric observations exceed significantly those of Doppler images 
obtained to date (see the most recent compilation of Doppler imaging results 
given by Strassmeier \cite{str1}). Doppler images, in turn, 
reveal key properties of spots, such as their latitudes, actual motion, 
areas, etc., which could strongly constrain conclusions made from light curve 
modeling. 

In this letter, using both techniques, we investigate a so-called 
``flip-flop'' phenomenon, which was discovered a decade ago in light curve 
behaviour of a single, late-type giant FK~Com (Jetsu et al.\ \cite{jetsu1},
\cite{jetsu2}, \cite{jetsu4}). In this phenomenon, the concentrated part of 
the spot activity shifts 180\degr\ in longitude over a short period of time 
and remains for some years at a new active longitude. Since its discovery, the 
``flip-flop'' phenomenon has also been reported in some RS~CVn binaries and 
in a single young dwarf LQ~Hya (Jetsu \cite{jetsu5}; Berdyugina \& Tuominen 
\cite{ber1}, \cite{ber5}; Berdyugina et al.\ \cite{ber2}, \cite{ber3}, 
\cite{ber4}; Rodon{\'o} et al.\ \cite{rodo}), implying that the phenomenon is 
a significant pattern of the activity in different types of stars and requires 
a thorough investigation.

The active, late-type, giant FK~Com itself is a prototype of a small group of 
rapidly rotating, single, giants (Bopp \& Stencel \cite{bopp3}). It exhibits 
strong and variable chromospheric and photospheric activity. Small variations 
in FK Com's visual magnitude were first reported by Chugainov (\cite{chu1}), 
and its spectral peculiarities include a strong and variable H$\alpha$ emission
 (e.g.\ Ramsey et al.\ \cite{ram}), strong chromospheric UV emission (Bopp \& 
Stencel \cite{bopp3}) and high X-ray luminosity (Walter \cite{wal}).

Since in FK~Com the ``flip-flop'' phenomenon was only observed as a shift of 
the light curve minimum, it was not clear whether the spots themselves move 
across the stellar surface. This could be only judged with the technique of 
Doppler imaging. The first hint came from the surface map for the year 1994 
(Korhonen et al.\ \cite{kor1}, from here-on Paper I), which seemed to present 
the star right at the moment of the ``flip-flop'', when one active region was 
getting weaker and another, in the opposite active longitude 180\degr\ apart, 
was taking over. Also, photometric observations have indicated that 
the two active longitudes in FK~Com can sometimes both be active at the same 
time (Jetsu et al.\ \cite{jetsu4}). This implies that the ``flip-flop'' in
FK~Com is caused by rearrangement of the spot activity at the 
active longitudes, and not by movement of the spots themselves. 

Peculiar phase shifts of light curve minima due to ``flip-flops''
being interpreted as movements of spots can result in variations of the 
spot rotation period determined by e.g.\ the Fourier analysis. Further, such 
period variations are normally interpreted as indications of the stellar 
differential rotation (e.g.\ Hall \cite{hall}; Henry et al.\ \cite{hen}), 
which can be therefore overestimated by few orders of 
magnitude. In FK~Com, for instance, an interval of the period variations is 
from 2$\fd$3960 to 2$\fd$4047 (Paper I) which could indicate the differential 
rotation rate larger than 0.004, while our Doppler imaging results limited it 
to 0.0001$\pm$0.0002, i.e.\ in fact gave evidence for a rigid rotation of the 
stellar surface (Korhonen et al.\ \cite{kor2}, from here-on Paper II).

The photometric behaviour of the other types of stars during the ``flip-flop'' 
phenomenon was found to be very similar to that of FK~Com. Moreover, Doppler 
imaging results for two  RS~CVn binaries II~Peg and IM~Peg clearly demonstrated
that in this type of stars the ``flip-flop'' is caused by diminishing spot area
in one active longitude and simultaneous increasing spot area in the
opposite longitude, i.e.\ switching the activity between the active longitudes 
due to relative spot area evolution rather than spot motion
(Berdyugina et al.\ \cite{ber2}, \cite{ber3}, \cite{ber4}). This suggests
that we could expect to observe something similar in FK~Com.

It was also noted that the flipping in RS~CVn stars happens periodically 
and can be predicted with a good accuracy (Berdyugina \& Tuominen \cite{ber1};
Rodon{\'o} et al.\ \cite{rodo}). In FK~Com, the full period of ``flip-flops'', 
i.e.\ when the activity returns to the same active longitude, was determined 
to be 6.5 years (Paper I). This period has been confirmed by the period 
analysis of 35 years of photometric observations (Korhonen et al.~\cite{kor3},
 \cite{kor4}, from here-on Papers III \& IV). The previous ``flip-flop'' has 
occurred during the summer 1994 (Papers I, III \& IV) and the next had to be 
expected in about 3.3 years, i.e.\ in the end of 1997. 

In this paper, we confirm the occurrence of the ``flip-flop'' in FK~Com
in the second half of 1997 and present the first temperature maps obtained 
right before and just after the ``flip-flop''. These maps clearly show that 
the ``flip-flop'' in FK~Com is caused by activity switching between the active 
longitudes without spot movement across the stellar surface, i.e.\ similarly 
to the RS~CVn stars.

\section{Observations}

Two sets of spectroscopic observations of FK~Com are used in this paper. The 
first set is from June 1997 and the details of these observations are given in 
Paper II. The second set was 
observed at the Kitt Peak National Observatory (KPNO) with the 0.9-m coud\'e 
feed telescope during 10 nights between 29th of December 1997 and 15th of 
January 1998. The TI-5 $800\times800$ CCD detector was employed together with 
grating~A, camera~5, the long collimator, and a 280-$\mu$m slit to give a 
resolving power of 38~000 at 6500 \AA. The useful wavelength range was 
80~\AA \ and the exposure time was set to 900~s. The reductions were done 
with the Image Reduction and Analysis Facility (IRAF) distributed by 
KPNO/NOAO. A summary of the KPNO observations can be found in 
Table~\ref{spect}. 

\begin{table}
\caption{Spectroscopic observations of FK~Com obtained between 29th of 
December 1997 and 15th of January 1998 at KPNO.}
\begin{tabular}{llllll}\hline
HJD    &Phase &S/N  &HJD     &Phase & S/N\\
2450000+&     &     &2450000+&      &    \\ 
\hline 
811.99 & 0.87 & 222 & 822.02 & 0.05 & 219 \\
814.00 & 0.71 & 291 & 823.03 & 0.47 & 273 \\
815.02 & 0.14 & 296 & 826.04 & 0.72 & 243 \\
820.02 & 0.22 & 218 & 828.04 & 0.55 & 252 \\
820.99 & 0.62 & 228 & 829.00 & 0.95 & 321 \\
\hline
\end{tabular}
\label{spect}
\end{table}

We also have simultaneous photometric observations in B and V from the 
Automatic Photometric Telescope (APT), Phoenix 10, Arizona, USA. For checking 
the reliability of the temperature map from January 1998 we have used 24 
observations taken between the 28th of November 1997 and the 7th of January 
1998. The observations obtained later than 7th of January are not included 
because of the very rapid change of the light curve in the beginning of 1998, 
especially for the earlier phases for which the spectroscopic observations 
have been obtained 8th of January or before. Observations with mean errors 
larger than $0\fm02$ were automatically excluded from the data set. These 
observations have been earlier published in Paper~III.

\section{The January 1998 temperature map}

In Paper~II, we found that during the years 1994--1997 the active region in 
FK~Com was shifting $0.22\pm 0.03$ in phase within a year. This phase shift 
was most likely caused by a difference between the used photometric period 
$2\fd4002466$ (from 25 year observations, Jetsu et al.\ \cite{jetsu2}) and the 
real spot rotation period for this time, $2\fd4037$ (Paper~II). The new map 
for January 1998 was first calculated with the old ephemeris given by Jetsu et
al.\ (\cite{jetsu2}): $2439252.895+2\fd4002466E$. Then, cross-correlation 
between the maps for June 1997 and January 1998 revealed that the rate of
$0.22\pm 0.03$ phase per year is still valid for the autumn 1997. To exclude 
the phase shift due to the period differences, the period $2\fd4037$ was 
used in this paper.

The inversions were done for all the lines in the spectral region between 
6416~{\AA} and 6444~{\AA} using the Tikhonov Regularization method code, 
INVERS7, written by N.\ Piskunov and modified by T.\ Hackman (Piskunov 
\cite{pisk1}, Hackman et al.\ \cite{hack}). The stellar parameters adopted for 
surface imaging can be found in Table~\ref{para}. More details on the method 
and the selection of the stellar parameters are given in Papers~I~\&~II. 
The model spectrum used in these calculations is the same as earlier used 
for the years 1996--1997 (Paper II).

\begin{table}
\caption{Adopted values of the stellar parameters for surface imaging.}\label{stellar}
\begin{tabular}{ll}\hline
Parameter & Adopted value\\ \hline
T$_{\rm eff}$ (unspotted)   & 5025~K\\
$\log g$                & 3.5 \\
Period                  & $2{\hbox{$.\!\!^{\rm d}$}}4037$ \\ 
$v\sin i$               & 155 \kms  \\
Inclination             & $60^{\circ}$ \\
Microturbulence         & 1.0 \kms \\ 
Macroturbulence         & 2.0 \kms \\ \hline
\end{tabular}
\label{para}
\end{table}

\begin{figure*}
\setlength{\unitlength}{1mm}
\begin{picture}(0,92)           
\put(10,-4){\begin{picture}(0,0) \includegraphics{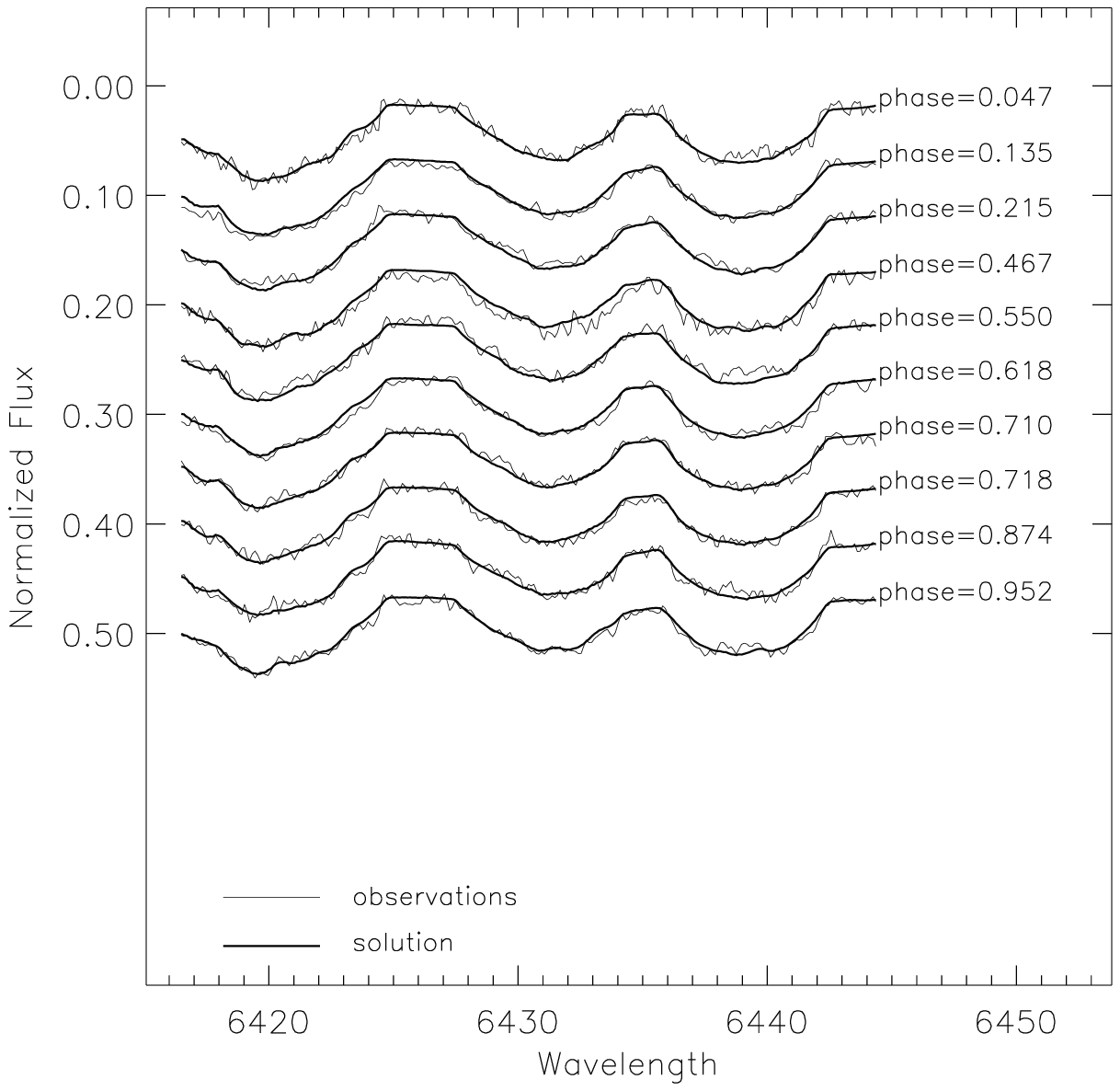} \end{picture}}
\put(65,-12){\begin{picture}(0,0) \includegraphics{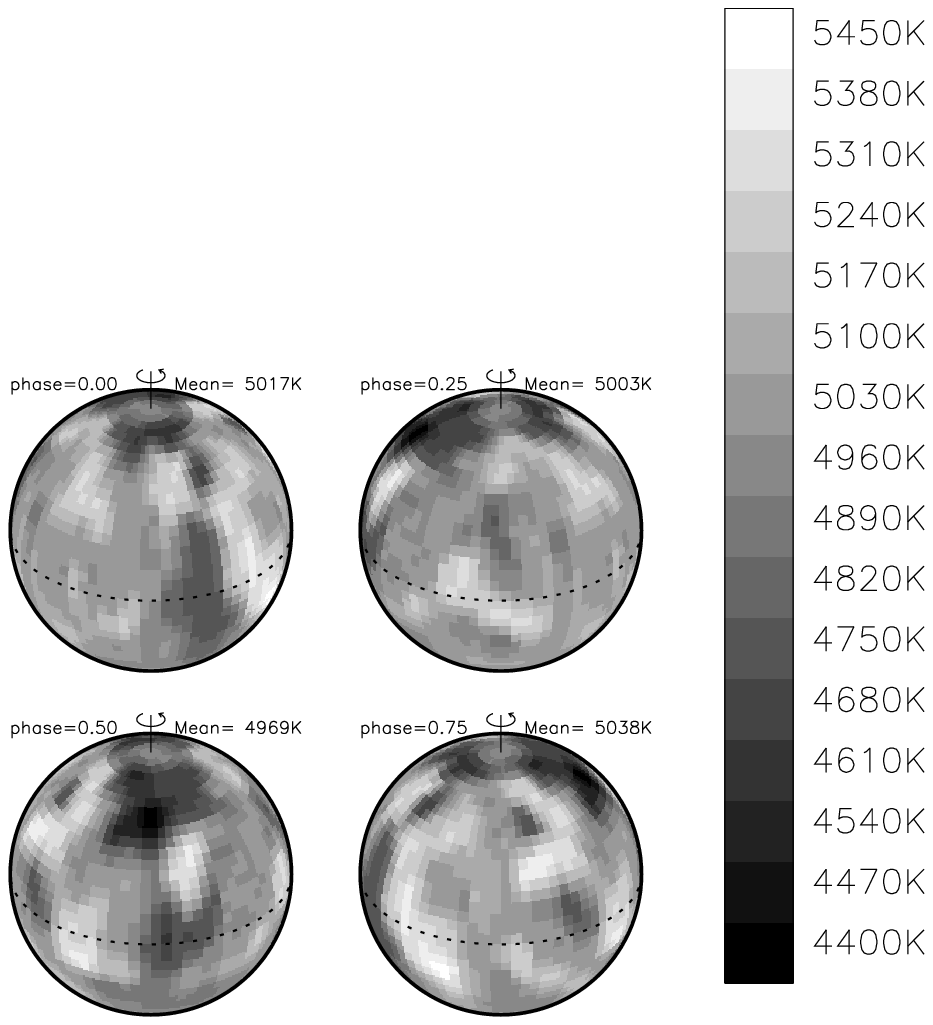} \end{picture}}
\put(93,60){\begin{picture}(0,0) \includegraphics{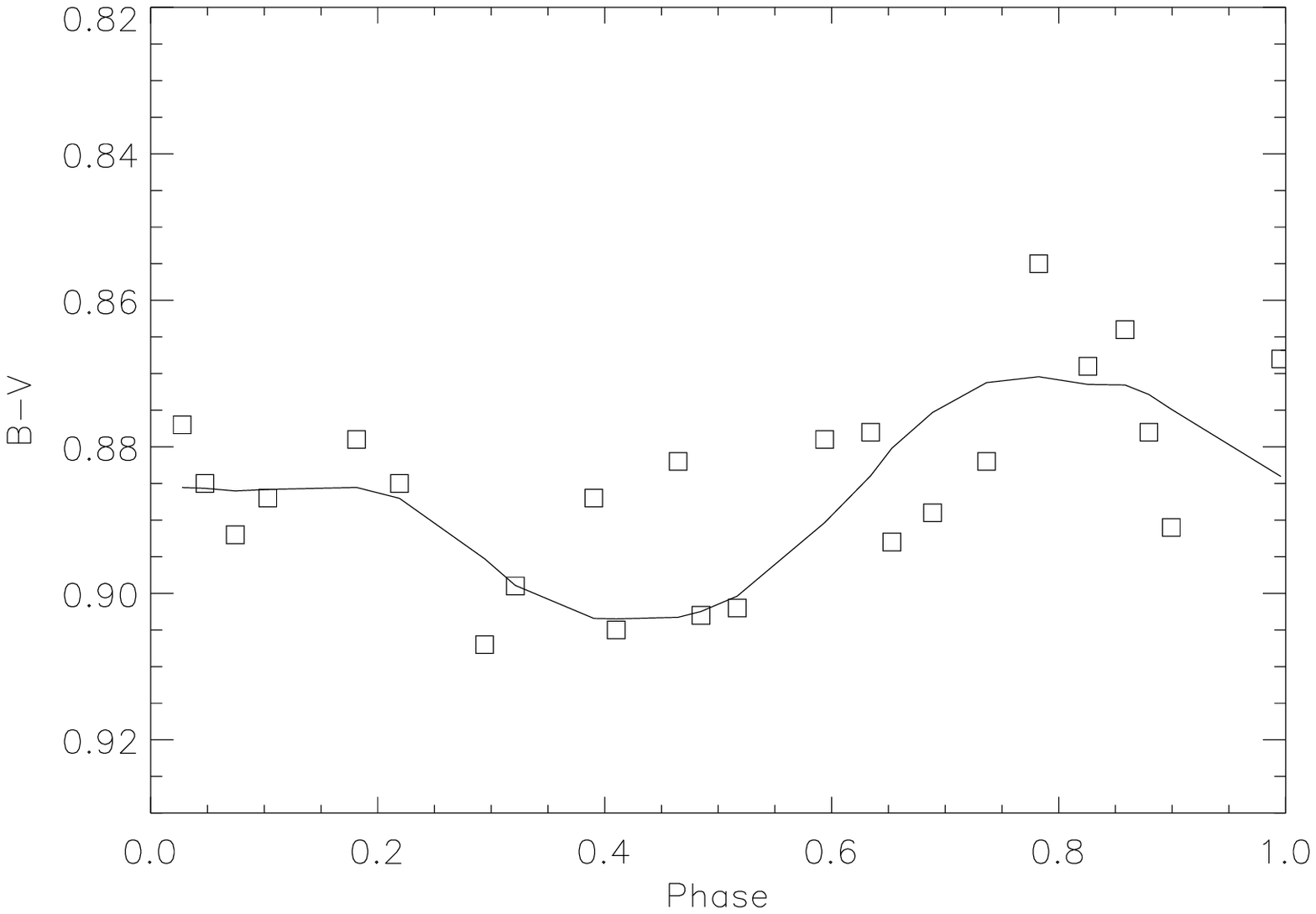} \end{picture}}
\put(20,60){\begin{picture}(0,0) \includegraphics{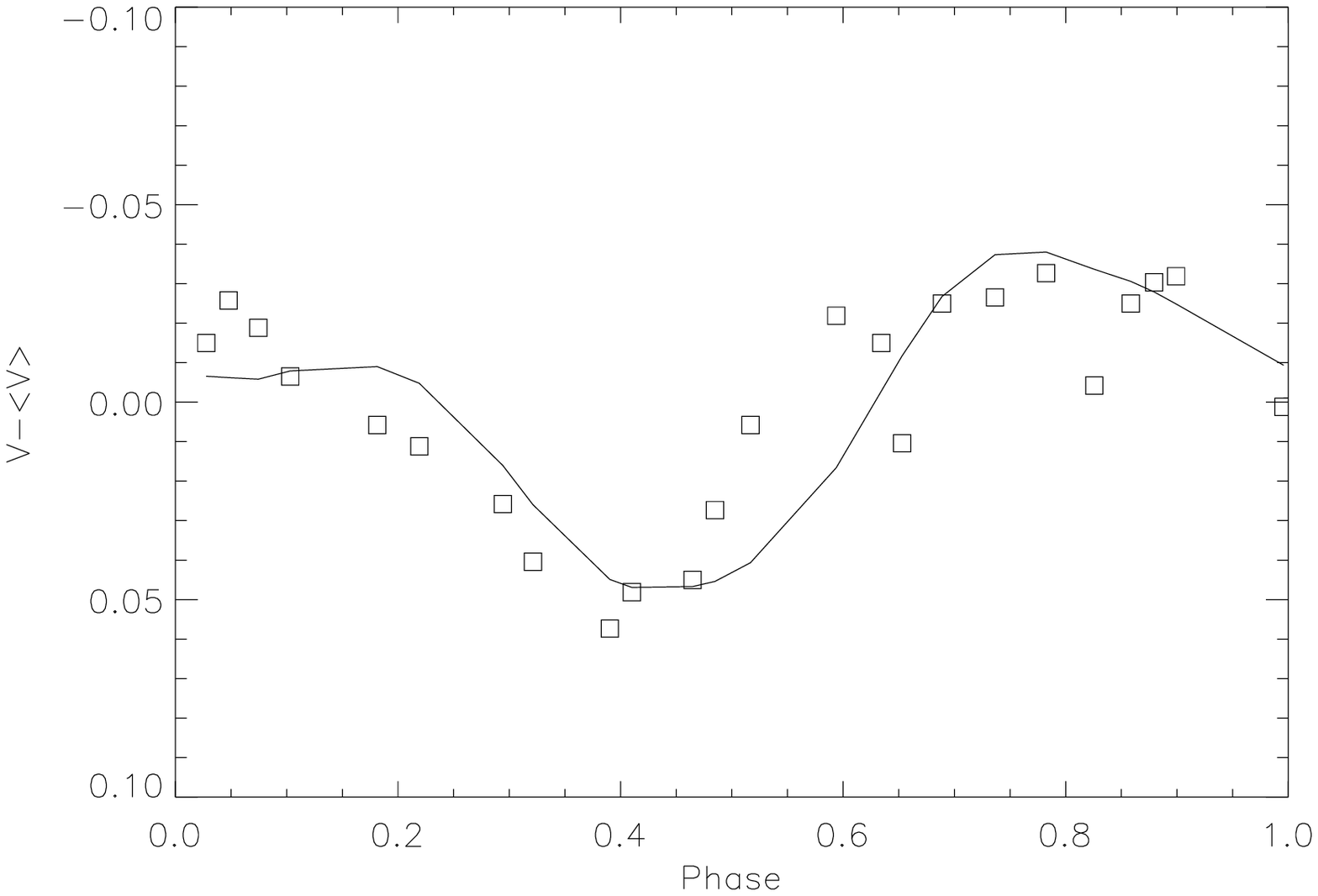} \end{picture}}
\end{picture}
\caption{The surface temperature map of FK~Com for December 1997 -- January 
1998 obtained with the Tikhonov regularization from the 6416--6444 {\AA} 
region with \vsini=155\kms. A grid of 40 latitudes and 80 longitudes across 
the stellar surface is used in the map. Calculated and observed spectral lines 
are shown by thick and thin lines, respectively. Photometric observations, from
 which the mean magnitude ($<{\rm V}>$) has been subtracted, are plotted with squares,
 and curves calculated from the map are presented by lines.}
\label{map97d}
\end{figure*}

The surface image and the fits to the observations are shown in 
Fig.~\ref{map97d}. The mean deviation of the spectroscopic observations from 
the model is 0.458~\%, which corresponds to an average S/N ratio of 218. 
The main feature is an extended cool area centered on phases $0.4\!-\!0.6$ 
in the latitude interval $45^{\circ}\!-\!72^{\circ}$. This feature has a 
temperature of $\sim\!600$~K below the unspotted surface. Another extended 
cool area with a temperature of $\sim\!500$~K below the unspotted temperature 
can be seen around the phases $0.8\!-\!1.1$ at approximately the same 
latitudes as the other cool area.

Photometric observations and the photometry calculated from the map are also 
shown in Fig.~\ref{map97d}. Note that they have not been used in the inversions
as a constrain. As can be seen, the B-V colours and the V magnitudes
calculated from the map fit the simultaneous photometric observations well. 
When looking at the V magnitude, the calculated curve seems to be very 
slightly shifted towards larger phases in comparison to the observations. 
Our photometry from the beginning of 1998 shows that the weaker active region 
at the phase $0.8\!-\!1.1$ evolves rapidly during January--March and becomes 
again stronger than it was in late 1997. The small shift in phase of the 
minimum of the calculated V magnitudes can most likely be explained by the 
fact that the spectroscopic observations were obtained over a time period of 
17 nights, covering 7.1 rotations, during a time of a rapid light curve 
evolution. Due to the reasonably good reproduction of the shape and position 
of the light curve minimum, the position and structure of the main features in 
the map are considered to be reliable. 

\section{The new ``flip-flop''}

\begin{figure}
\setlength{\unitlength}{1mm}          
\begin{picture}(0,26) 
\put(1,0){\begin{picture}(0,0) \includegraphics{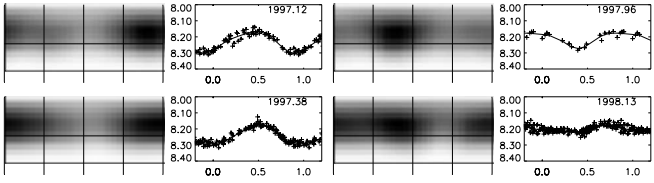} \end{picture}}
\end{picture}
\caption{The results of the light curve inversions for the year 1997 and 
beginning of the year 1998. The darker regions indicate larger spot filling 
factor. The grid in the maps indicates the equator and 4 longitudes separated 
by 90\degr. In the photometry the observations are presented by crosses and 
the calculated curve with a line.}
\label{inv}
\end{figure}

The photometric observations from the 1990's (Papers III \& IV) show that 
the ``flip-flops'' have occurred during the summer 
1994 and again during autumn 1997, in accordance with the 3.3-year cycle. 
In Fig.~\ref{inv} the photometric observations for 1997 and beginning 
of 1998 are shown together with the spot filling factor maps obtained by 
inversions from the light curves. The light curve inversions are 
done using a technique similar to that used by Rodon{\'o} 
et al.~(\cite{rodo}) and explained in more detail in Paper IV 
where the results of inversions for 35 years of photometric 
observations of FK~Com are presented.

Fig.~\ref{inv} clearly shows the ``flip-flop'' that occurred during the autumn
1997. In the second half of 1997, the old active region around the phase 1.0
has significantly diminished, while the new one near the phase 0.5 has 
appeared. In the beginning of 1998, the old active region becomes again 
rapidly stronger than it was during December 1997. The photometry for 1998--1999 
(Papers III \& IV) shows that from January 1998 till June 1999 both active 
regions are simultaneously active, resulting in a very flat and featureless 
light curve.

\begin{figure}
\setlength{\unitlength}{1mm}          
\begin{picture}(0,20)
\put(0,0){\begin{picture}(0,0) \includegraphics{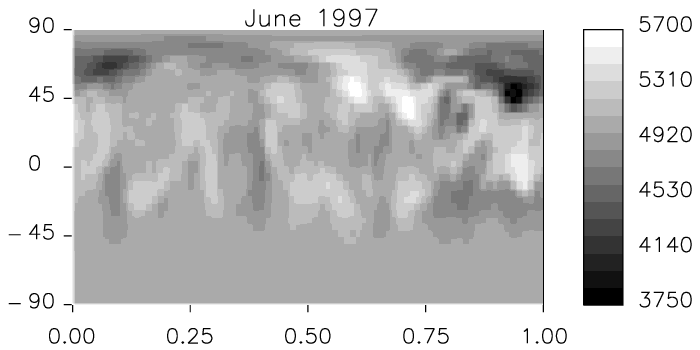} \end{picture}}
\put(45,0){\begin{picture}(0,0) \includegraphics{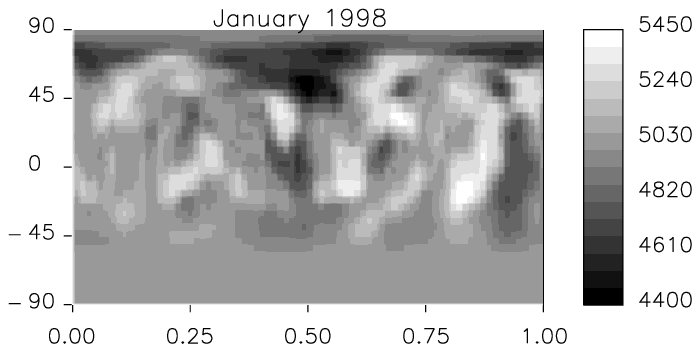} \end{picture}}
\end{picture}
\caption{The surface temperature maps for June 1997 and January 1998. 
The grey scale gives the temperatures in Kelvin.}
\label{maps}
\end{figure}

Fig.~\ref{maps} shows the temperature maps obtained with surface imaging for 
June 1997 and January 1998. In June 1997, the spots are concentrated 
around the phase 1.0 and there is no evidence of spots at the phase 0.5. 
In January 1998, the most contrast spot group is seen at the phase 0.5 and 
a secondary group is around the phase 1.0. It is reasonable to assume that 
the spot group at the phase 1.0 during January 1998 is the same as the one 
in the June 1997 map at the same phase, the activity of the spot group has 
just diminished. The spot group, 0.5 in phase apart from the old active region,
is a new active region which has appeared on the other side of the star, as 
predicted in the ``flip-flop'' phenomenon. 
The cross-correlation of the maps, done as described in Paper II,
using latitudes 42$\fdg$75--83$\fdg$25, yields phase shifts of $-0.04$ 
and $0.50$ for the old and new active regions, respectively.

In FK~Com, usually only one of the active longitudes is active at a time, 
as can be seen, for example, from the temperature maps for the years 
1995--1997 (Papers~I~\&~II). Anyhow, near the ``flip-flop'' both active 
longitudes become simultaneously active, like in summer 1994 (Paper~I)
and in 1997--1998 (Fig.~\ref{inv}), and the change of the relative strengths 
of the active regions causes the ``flip-flop''.

\section{Conclusions}

With the Doppler imaging technique and light curve inversions we showed clearly
 that FK~Com possesses the same spot activity pattern as observed in binary 
components of the RS~CVn type and in a young dwarf LQ~Hya, namely 
{\it the ``flip-flop'' phenomenon caused by changing the relative strengths 
of the spot groups at the two active longitudes}. The fact that different 
types of rapidly rotating stars show the same phenomenon in detail supports 
the idea that it is the rapid rotation that determines the principal component 
of the stellar activity, {\it regardless of the origin of the rapid rotation}.
 
Since during the ``flip-flop'' the light curve undergoes rapid and irregular
changes, one should be careful in interpreting such changes as spot period
variations and, as a consequence, the stellar differential rotation.

\begin{acknowledgements}
The work of HK was supported by the Finnish graduate school in Astronomy and 
Space Physics and Vilho, Yrj{\"o} and Kalle V{\"a}is{\"a}l{\"a} Foundation. 
KGS acknowledges the receipt of FWF grant S7301 from the Austrian Science 
Foundation.
\end{acknowledgements}

\end{document}